# Relativistic shape invariant potentials

A. D. Alhaidari

Physics Department, King Fahd University of Petroleum & Minerals, Box 5047, Dhahran 31261, Saudi Arabia

E-mail: **haidari@mailaps.org**

Dirac equation for a charged spinor in electromagnetic field is written for special cases of spherically symmetric potentials. This facilitates the introduction of relativistic extensions of shape invariant potential classes. We obtain the relativistic spectra and spinor wavefunctions for all potentials in one of these classes. The nonrelativistic limit reproduces the usual Rosen-Mörse I & II, Eckart, Pöschl-Teller, and Scarf potentials.

Exactly solvable potentials in nonrelativistic quantum mechanics fall within distinct classes of "shape invariant potentials" [1–8]. Each potential in a given class can be mapped into another in the same class by a canonical transformation of the coordinates [7,9–13]. The transformation gives a correspondence map among the potential parameters, angular momentum, and energy. Using the resulting parameter substitution and the bound states spectrum of the original potential one can easily and directly obtain the spectra of all other potentials in the class. Moreover, the eigenstates wavefunctions are obtained by simple transformations of those of the original potential. It is very tempting to search for the relativistic extensions of these classes and obtain the relativistic spectra of the bound states and corresponding spinor wavefunctions. In fact, one such class has already been established. Recently, the Dirac-Mörse potential was introduced and its bound states spectrum and spinor wavefunctions were obtained [14]. Together with its two well-established partners, the Dirac-Coulomb and Dirac-Oscillator [15] potentials, they complete one relativistic class. In this letter, we continue these efforts by introducing the relativistic extension of yet another class of shape invariant potentials that includes "Dirac-Rosen-Mörse", "Dirac-Eckart", "Dirac-Pöschl-Teller", and "Dirac-Scarf" potentials. We obtain their relativistic bound states spectra and spinor wavefunctions. This is accomplished by following the same procedure that was used in reference [14] for the introduction and solution of the Dirac-Mörse problem.

We start by setting up the physical problem of a charged spinor in spherically symmetric four-component electromagnetic potential. Gauge invariance and spherically symmetry of the electrostatic potential is used to arrive at the radial Dirac equation. Afterwards, we apply a unitary transformation to Dirac equation such that the resulting second order differential equation becomes Schrödinger-like so that comparison with exactly solvable nonrelativistic problems is transparent. Thus, the resulting simple correspondence map among parameters of the two problems gives the sought after bound states spectrum and wavefunction.

In atomic units ($m = e = \hbar = 1$) and taking the speed of light $c = \alpha^{-1}$, the Hamiltonian for a Dirac spinor in four-component electromagnetic potential, $(A_0, \vec{A})$, reads

$$H = \begin{pmatrix} 1+\alpha A_0 & -i\alpha\vec{\sigma}\cdot\vec{\nabla} + \alpha\vec{\sigma}\cdot\vec{A} \\ -i\alpha\vec{\sigma}\cdot\vec{\nabla} + \alpha\vec{\sigma}\cdot\vec{A} & -1+\alpha A_0 \end{pmatrix}$$

where $\alpha$ is the fine structure constant and $\vec{\sigma}$ are the three 2×2 Pauli spin matrices. In quantum electrodynamics (the theory of interaction of charged particles with the electromagnetic field), local gauge symmetry implies invariance under the transformation

$$\left(A_0, \vec{A}\right) \to \left(A_0, \vec{A}\right) + \left(\alpha \partial \Lambda/\partial t, \vec{\nabla} \Lambda\right)$$

where $\Lambda(t,\vec{r})$ is a real space-time function. That is adding a 4-dimensional gradient of the gauge field $\Lambda(t,\vec{r})$ to the electromagnetic potential will not alter the physical content of the theory. In the lab frame, gauge invariance implies that the general form of the electromagnetic potential for <u>static</u> charge distribution with <u>spherical</u> symmetry is

$$\left(A_0, \vec{A}\right) = \left(\alpha V(r), \vec{0}\right) + \left(0, \vec{\nabla} \Lambda(r)\right) \equiv \left(\alpha V(r), \hat{r} W(r)\right)$$

where $V(r)$ is the electrostatic potential function and $\hat{r}$ is the radial unit vector. Obviously, $W(r)$ is a gauge field that does not contribute to the magnetic field. However, fixing this gauge degree of freedom by taking $W = 0$ is not the best choice. An alternative and proper "gauge fixing condition", which is much more fruitful, will be imposed as a constraint in equation (4) below. We will consider, however, an alternative coupling of the electromagnetic potential to the charged Dirac particle. The two off-diagonal terms $\alpha \vec{\sigma} \cdot \vec{A}$ in the Hamiltonian $H$ above are to be replaced by $\pm i\alpha \vec{\sigma} \cdot \vec{A}$, respectively, resulting in the following two-component radial Dirac equation

$$\begin{pmatrix} 1 + \alpha^2 V(r) & \alpha\left(\dfrac{\kappa}{r} + W(r) - \dfrac{d}{dr}\right) \\ \alpha\left(\dfrac{\kappa}{r} + W(r) + \dfrac{d}{dr}\right) & -1 + \alpha^2 V(r) \end{pmatrix} \begin{pmatrix} g(r) \\ f(r) \end{pmatrix} = \varepsilon \begin{pmatrix} g(r) \\ f(r) \end{pmatrix} \qquad (1)$$

where $\varepsilon$ is the relativistic energy and $\kappa$ is the spin-orbit coupling parameter defined as $\kappa = \pm (j + \frac{1}{2})$ for $l = j \pm \frac{1}{2}$. Equation (1) gives two coupled first order differential equations for the two radial spinor components. By eliminating the lower component we obtain a second order differential equation for the upper. The resulting equation may turn out to be not Schrödinger-like, i.e. it may contain first order derivatives. We apply a general local unitary transformation that eliminates the first order derivative as follows:

$$r = q(x) \quad \text{and} \quad \begin{pmatrix} g(r) \\ f(r) \end{pmatrix} = \begin{pmatrix} \cos(\rho(x)) & \sin(\rho(x)) \\ -\sin(\rho(x)) & \cos(\rho(x)) \end{pmatrix} \begin{pmatrix} \phi(x) \\ \theta(x) \end{pmatrix} \qquad (2)$$

The stated requirement gives the following constraint:

$$\frac{dq}{dx}\left[-\alpha^2 V + \cos(2\rho) + \alpha \sin(2\rho)(W + \kappa/q) + \alpha \frac{d\rho/dx}{dq/dx} + \varepsilon\right] = \text{constant} \equiv \eta \neq 0 \qquad (3)$$

This transformation and the resulting constraint are the relativistic analog of point canonical transformation in nonrelativistic quantum mechanics [7,9–13]. In this article, we consider the case of global unitary transformation defined by $q(x) = x$ and $d\rho/dx = 0$. Substituting these in the constraint equation (3) yields

$$W(r) = \frac{\alpha}{S} V(r) - \frac{\kappa}{r}$$
$$\eta = C + \varepsilon \qquad (4)$$

where $S \equiv \sin(2\rho)$ and $C \equiv \cos(2\rho)$. The first relation in (4) is the gauge fixing condition for the electromagnetic potential. The transformation defined above subject to the constraint maps the radial Dirac equation (1) into the following:

$$\begin{pmatrix} C+2\alpha^2 V & \alpha\left(-\dfrac{S}{\alpha}+\dfrac{\alpha C}{S}V-\dfrac{d}{dr}\right) \\ \alpha\left(-\dfrac{S}{\alpha}+\dfrac{\alpha C}{S}V+\dfrac{d}{dr}\right) & -C \end{pmatrix}\begin{pmatrix}\phi(r)\\ \theta(r)\end{pmatrix} = \varepsilon\begin{pmatrix}\phi(r)\\ \theta(r)\end{pmatrix}$$

which in turn gives an equation for the lower spinor component in terms of the upper:

$$\theta(r) = \frac{\alpha}{C+\varepsilon}\left[-\frac{S}{\alpha}+\frac{\alpha C}{S}V+\frac{d}{dr}\right]\phi(r) \tag{5}$$

Giving the following Schrödinger-like 2$^{nd}$ order differential equation for the upper component:

$$\left[-\frac{d^2}{dr^2}+\frac{\alpha^2}{T^2}V^2+2\varepsilon V-\frac{\alpha}{T}\frac{dV}{dr}-\frac{\varepsilon^2-1}{\alpha^2}\right]\phi(r)=0 \tag{6}$$

where $T \equiv S/C = \tan(2\rho)$.

Nonrelativistic shape invariant potentials can be divided in two classes based on the form of their eigenfunctions. In the first class, which includes the Coulomb, Oscillator and Mörse potentials, the wavefunctions are written in terms of the Confluent hypergeometric functions. The relativistic extension of this class has already been established [14,15]. In the second class, which is of interest to our present investigation, the wavefunctions are written in terms of the hypergeometric functions. This class includes Rosen-Mörse, Eckart, Pöschl-Teller, and Scarf potentials. The algebraic expressions of these potentials and their properties are given in [3-5,7,8] and references therein. Specifically, we will consider the hyperbolic rather than the trigonometric form of these potentials. Therefore, in our attempt to search for the relativistic extension of these potentials we will consider expressions for $V(r)$ or $W(r)$ which are simple linear combinations of sinh($r$), sech($r$), tanh($r$), etc. such that the nonrelativistic potentials are reproduced in the limit. Our use of the terms "simple" and "linear" in the previous statement is due to the fact that these are dominant properties of the relativistic theory. As examples: (1) Dirac equation is linear in the derivative whereas Schrödinger equation is quadratic; (2) the Dirac-Oscillator potential [15] is linear in the coordinate while the oscillator potential is quadratic; (3) the Dirac-Mörse potential [14] is linear in the exponential (i.e. of the form $e^{-x}$) whereas the nonrelativistic Mörse potential is of the form $(1-e^{-x})^2$.

Now, let us consider the case where the potential function $V(r) = D\tanh(\lambda r)$ with $D$ and $\lambda$ being real parameters. Equation (6) gives the following second order differential equation for the upper spinor component

$$\left[-\frac{d^2}{dr^2}-\frac{\alpha D}{T}\left(\frac{\alpha D}{T}+\lambda\right)\frac{1}{\cosh^2(\lambda r)}+2\varepsilon D\tanh(\lambda r)+\left(\frac{\alpha D}{T}\right)^2-\frac{\varepsilon^2-1}{\alpha^2}\right]\phi(r)=0$$

We compare this with Schrödinger equation for the S-wave Rosen-Mörse I potential [7]

$$\left[-\frac{d^2}{dr^2}-A(A+\lambda)\frac{1}{\cosh^2(\lambda r)}+2B\tanh(\lambda r)+A^2-2E\right]\phi(r)=0 \tag{7}$$

where $A$, $B$, and $\lambda$ are real constant parameters with $\lambda A > 0$, and $E$ is the nonrelativistic energy. The comparison gives the following correspondence between nonrelativistic and relativistic parameters:

$$A = \alpha D/T$$
$$B = D\varepsilon \qquad (8)$$
$$E = (\varepsilon^2 - 1)/2\alpha^2$$

The well-known nonrelativistic bound states spectrum of equation (7) is

$$E_n = -\frac{\lambda^2}{2}(A/\lambda - n)^2 - \frac{(B/\lambda)^2}{2(A/\lambda - n)^2} + \frac{A^2}{2}, \quad n = 0,1,\ldots,n_{\max} < A/\lambda \qquad (9)$$

The substitution formulas in (8) give the following spectrum for this relativistic "Dirac-Rosen-Mörse I" potential

$$\varepsilon_n = \left\{ \left[ 1 + (\alpha^2 D/T)^2 - \frac{\alpha^2 \lambda^2}{(\alpha D/\lambda T - n)^2} \right] \Big/ \left[ 1 + \left( \frac{\alpha D/\lambda}{\alpha D/\lambda T - n} \right)^2 \right] \right\}^{1/2}$$

where $n = 0,1,2,\ldots,n_{\max}$ and $n_{\max}$ is the smallest integer satisfying

$$\left| n_{\max} - \frac{\alpha D}{\lambda T} \right| > \frac{1}{\sqrt{(\alpha D/\lambda T)^2 + (\alpha\lambda)^{-2}}}$$

Taking the nonrelativistic limit of this spectrum with

$$\alpha \to 0$$
$$\varepsilon_n \approx 1 + \alpha^2 E_n$$
$$T \approx \alpha\tau$$

reproduces the nonrelativistic spectrum (9) with $\tau = D/A$. The bound states wavefunction of the nonrelativistic problem [7] is mapped, using (8), into the following upper spinor component wavefunction

$$\phi_n(r) = R_n (1-z)^{(\beta-n+\gamma_n)/2} (1+z)^{(\beta-n-\gamma_n)/2} P_n^{(\beta-n+\gamma_n,\beta-n-\gamma_n)}(z)$$

where $P_n^{(\mu,\nu)}(z)$ is the Jacobi polynomial [16], $R_n$ is the normalization constant, and

$$z = \tanh(\lambda r)$$
$$\beta = \alpha D/\lambda T$$
$$\gamma_n = \frac{D\varepsilon_n/\lambda^2}{\beta - n}$$

Equation (5) gives the lower spinor component in terms of the upper as

$$\theta_n(r) = \frac{\alpha\lambda}{\varepsilon_n + C}\left[ -\frac{S}{\alpha\lambda} + \beta z + (1-z^2)\frac{d}{dz} \right]\phi_n(r)$$

Using the differential and recursion properties of the Jacobi polynomials [16], we can write this explicitly as

$$\theta_n(r) = \frac{\alpha\lambda/\beta}{\varepsilon_n + C} R_n (1-z)^{\mu/2}(1+z)^{\nu/2}\left[ -(D/\lambda^2)(\varepsilon_n + C)P_n^{(\mu,\nu)}(z) + (\beta^2 - \gamma_n^2)P_{n-1}^{(\mu,\nu)}(z) \right]$$

where $\mu = \beta - n + \gamma_n$ and $\nu = \beta - n - \gamma_n$.

If we now take the alternative choice of potential, $V(r) = -D\coth(\lambda r)$, and go through the same steps above we arrive at the relativistic extension of Eckart potential [3,7]. The bound states spectrum and spinor wavefunction for this relativistic "Dirac-Eckart" potential are listed in the Table.

To obtain the relativistic extension of the other potentials in this class we consider the case $V = 0$ which is equivalent to the identity transformation (i.e. $\rho = 0$) combined with the constraint (4). Thus, Dirac equation (1) now reads

$$\begin{pmatrix} 1 & \alpha\left(W + \dfrac{\kappa}{r} - \dfrac{d}{dr}\right) \\ \alpha\left(W + \dfrac{\kappa}{r} + \dfrac{d}{dr}\right) & -1 \end{pmatrix} \begin{pmatrix} \phi(r) \\ \theta(r) \end{pmatrix} = \varepsilon \begin{pmatrix} \phi(r) \\ \theta(r) \end{pmatrix}$$

This gives the following equation for the lower spinor component in terms of the upper:

$$\theta(r) = \frac{\alpha}{1+\varepsilon}\left(W + \frac{\kappa}{r} + \frac{d}{dr}\right)\phi(r) \tag{10}$$

While, the upper component solves the following Schrödinger-like 2$^{nd}$ order differential equation

$$\left[-\frac{d^2}{dr^2} + \frac{\kappa(\kappa+1)}{r^2} + W^2 - \frac{dW}{dr} + 2\kappa\frac{W}{r} - \frac{\varepsilon^2-1}{\alpha^2}\right]\phi(r) = 0 \tag{11}$$

Now, all nonrelativistic potentials in this class are solvable only for the S-wave problem (i.e. $l = 0$), thus we restrict our analysis to the case where $\kappa = 0$. We start by considering $W(r) = F \coth(\lambda r) - G \operatorname{csch}(\lambda r)$ with $F$, $G$, and $\lambda$ being real constant parameters and $\lambda F > 0$. With this potential function and $\kappa = 0$, equation (11) gives the following second order differential equation for the upper spinor component

$$\left[-\frac{d^2}{dr^2} + \frac{F^2 + G^2 + \lambda F}{\sinh^2(\lambda r)} - G(2F+\lambda)\frac{\cosh(\lambda r)}{\sinh^2(\lambda r)} + F^2 - \frac{\varepsilon^2-1}{\alpha^2}\right]\phi(r) = 0$$

Comparing this with Schrödinger equation for the S-wave Rosen-Morse II potential [3,7]

$$\left[-\frac{d^2}{dr^2} + \frac{A^2 + B^2 + \lambda A}{\sinh^2(\lambda r)} - B(2A+\lambda)\frac{\cosh(\lambda r)}{\sinh^2(\lambda r)} + A^2 - 2E\right]\phi(r) = 0 \tag{12}$$

gives the following correspondence between nonrelativistic and relativistic parameters:

$$A = F$$
$$B = G \tag{13}$$
$$E = (\varepsilon^2 - 1)/2\alpha^2$$

The well-known nonrelativistic bound states spectrum of equation (12) is

$$E_n = -\frac{\lambda^2}{2}(A/\lambda - n)^2 + \frac{A^2}{2}, \quad n = 0, 1, \ldots, n_{\max} < A/\lambda \tag{14}$$

The substitution (13) results in the following relativistic spectrum for this "Dirac-Rosen-Mörse II" potential

$$\varepsilon_n = \pm\sqrt{1 + \alpha^2 F^2 - \alpha^2 \lambda^2 (F/\lambda - n)^2} \tag{15}$$

where $n = 0, 1, 2, \ldots, n_{\max}$ and $n_{\max}$ is the largest integer satisfying

$$|n_{\max} - F/\lambda| < \sqrt{(F/\lambda)^2 + (\alpha\lambda)^{-2}}$$

It is obvious that the nonrelativistic limit ($\alpha \to 0$) of (15) reproduces the spectrum in (14). The bound states wavefunction of the nonrelativistic problem [3,7] is transformed, using (13), into the following upper spinor component wavefunction

$$\phi_n(r) = R_n (z-1)^{(\gamma-\beta)/2}(z+1)^{-(\gamma+\beta)/2} P_n^{(\gamma-\beta-1/2,-\gamma-\beta-1/2)}(z)$$

where

$$z = \cosh(\lambda r)$$
$$\beta = F/\lambda$$
$$\gamma = G/\lambda$$

Equation (10) gives the lower spinor component in terms of the upper as

$$\theta_n(r) = -\frac{\alpha\lambda}{\varepsilon_n + 1}(z^2 - 1)^{-1/2}\left[\gamma - \beta z + (1 - z^2)\frac{d}{dz}\right]\phi_n(r)$$

Again, using the differential and recursion properties of the Jacobi polynomials [16], we can write this explicitly as

$$\theta_n(r) = \frac{\alpha\lambda}{\varepsilon_n + 1} R_n(z-1)^{(\mu-\frac{1}{2})/2}(z+1)^{(\nu-\frac{1}{2})/2}\left\{n\left(z + \frac{\gamma}{\beta - n + 1/2}\right)P_n^{(\mu,\nu)}(z)\right.$$
$$\left. + \left[\frac{(\beta - n + \frac{1}{2})^2 - \gamma^2}{\beta - n + 1/2}\right]P_{n-1}^{(\mu,\nu)}(z)\right\}$$

where $\mu = \gamma - \beta - 1/2$ and $\nu = -\gamma - \beta - 1/2$.

Taking the alternative choice $W(r) = F\tanh(\lambda r) + G\operatorname{sech}(\lambda r)$ and going through the same steps above we arrive at the relativistic extension of Scarf potential [3,17]. The bound states spectrum and spinor wavefunction for this "Dirac-Scarf" potential are listed in the Table. The table also lists "Dirac-Pöschl-Teller" potential $W(r) = F\tanh(\lambda r) - G\coth(\lambda r)$ which, in the nonrelativistic limit, reproduces the usual Pöschl-Teller potential [7,18,19].

Finally, it is worth noting that it would be of prime relevance, as a future development, to find the general transformations $q(x)$ and $\rho(x)$ in (2) that map any one of these relativistic potentials into other members of the class. Moreover, it might be possible that an exhaustive study of such transformations may bring about new relativistic potentials that enlarge the class. A similar treatment is called for concerning the other class of relativistic potentials that includes Dirac-Coulomb, Dirac-Oscillator, and Dirac-Mörse potentials.

**Acknowledgements:** The author is grateful to Dr. M. S. Abdelmonem for the invaluable support in literature search.

**Table Caption:**

Potential functions $V(r)$ and $W(r)$, transformation angle $\rho$, and bound states spectrum $\varepsilon_n$ for the five potentials. The table is continued to show explicitly the two-component radial spinor wavefunctions $\phi_n(r)$ and $\theta_n(r)$ for each potential.

**Table**

|  | $V(r)$ | $W(r)$ | $\tan(2\rho)$ | $\varepsilon_n$ |
|---|---|---|---|---|
| **Dirac-Rosen-Mörse I** | $D\tanh(\lambda r)$ | $(\alpha D/S)\tanh(\lambda r) - \kappa/r$ | $\alpha D/A$ | $\left\{\left[1+\left(\alpha^2 D/T\right)^2 - \dfrac{\alpha^2\lambda^2}{(\alpha D/\lambda T - n)^2}\right]\Big/\left[1+\left(\dfrac{\alpha D/\lambda}{\alpha D/\lambda T - n}\right)^2\right]\right\}^{1/2}$ |
| **Dirac-Eckart** | $-D\coth(\lambda r)$ | $-(\alpha D/S)\coth(\lambda r) - \kappa/r$ | $\alpha D/A$ | $\left\{\left[1+\left(\alpha^2 D/T\right)^2 - \dfrac{\alpha^2\lambda^2}{(\alpha D/\lambda T + n)^2}\right]\Big/\left[1+\left(\dfrac{\alpha D/\lambda}{\alpha D/\lambda T + n}\right)^2\right]\right\}^{1/2}$ |
| **Dirac-Rosen-Mörse II** | $0$ | $F\coth(\lambda r) - G\operatorname{csch}(\lambda r)$ | $0$ | $\sqrt{1+\alpha^2 F^2 - \alpha^2\lambda^2\left(F/\lambda - n\right)^2}$ |
| **Dirac-Scarf** | $0$ | $F\tanh(\lambda r) + G\operatorname{sech}(\lambda r)$ | $0$ | $\sqrt{1+\alpha^2 F^2 - \alpha^2\lambda^2\left(F/\lambda - n\right)^2}$ |
| **Dirac-Pöschl-Teller** | $0$ | $F\tanh(\lambda r) - G\coth(\lambda r)$ | $0$ | $\sqrt{1+\alpha^2(G-F)^2 - \alpha^2\lambda^2\left[(G-F)/\lambda + 2n\right]^2}$ |

**Table (continued)**

| | $\phi_n(r)$ | $\theta_n(r)$ |
|---|---|---|
| **Dirac-Rosen-Mörse I** | $\phi_n(r) = R_n(1-z)^{\mu/2}(1+z)^{\nu/2} P_n^{(\mu,\nu)}(z)$<br>$z = \tanh(\lambda r), \mu = \beta - n + \gamma_n, \nu = \beta - n - \gamma_n$<br>$\beta = \alpha D/\lambda T, \gamma_n = (\varepsilon_n D/\lambda^2)(\beta - n)^{-1}$ | $\theta_n(r) = \dfrac{\alpha\lambda/\beta}{\varepsilon_n + C} R_n(1-z)^{\mu/2}(1+z)^{\nu/2} \times$<br>$\left[ -(D/\lambda^2)(\varepsilon_n + C) P_n^{(\mu,\nu)}(z) + (\beta^2 - \gamma_n^2) P_{n-1}^{(\mu,\nu)}(z) \right]$ |
| **Dirac-Eckart** | $\phi_n(r) = R_n(z-1)^{\mu/2}(z+1)^{\nu/2} P_n^{(\mu,\nu)}(z)$<br>$z = \coth(\lambda r), \mu = -\beta - n + \gamma_n, \nu = -\beta - n - \gamma_n$<br>$\beta = \alpha D/\lambda T, \gamma_n = (\varepsilon_n D/\lambda^2)(\beta + n)^{-1}$ | $\theta_n(r) = \dfrac{\alpha\lambda/\beta}{\varepsilon_n + C} R_n(z-1)^{\mu/2}(z+1)^{\nu/2} \times$<br>$\left[ -(D/\lambda^2)(\varepsilon_n + C) P_n^{(\mu,\nu)}(z) + (\gamma_n^2 - \beta^2) P_{n-1}^{(\mu,\nu)}(z) \right]$ |
| **Dirac-Rosen-Mörse II** | $\phi_n(r) = R_n(z-1)^{(\mu+\frac{1}{2})/2}(z+1)^{(\nu+\frac{1}{2})/2} P_n^{(\mu,\nu)}(z)$<br>$z = \cosh(\lambda r), \beta = F/\lambda, \gamma = G/\lambda$<br>$\mu = -\beta - \tfrac{1}{2} + \gamma, \nu = -\beta - \tfrac{1}{2} - \gamma$ | $\theta_n(r) = \dfrac{\alpha\lambda}{\varepsilon_n + 1} R_n (z-1)^{\frac{\mu-\frac{1}{2}}{2}}(z+1)^{\frac{\nu-\frac{1}{2}}{2}} \left\{ n\left( z + \dfrac{\gamma}{\beta - n + 1/2} \right) P_n^{(\mu,\nu)}(z) \right.$<br>$\left. + \left[ \dfrac{(\beta - n + \frac{1}{2})^2 - \gamma^2}{\beta - n + 1/2} \right] P_{n-1}^{(\mu,\nu)}(z) \right\}$ |
| **Dirac-Scarf** | $\phi_n(r) = R_n(1+z^2)^{-\beta/2} e^{-\gamma \tan^{-1}(z)} P_n^{(\mu,\nu)}(iz)$<br>$z = \sinh(\lambda r), \beta = F/\lambda, \gamma = G/\lambda$<br>$\mu = -\beta - \tfrac{1}{2} - i\gamma, \nu = -\beta - \tfrac{1}{2} + i\gamma$ | $\theta_n(r) = \dfrac{\alpha\lambda}{\varepsilon_n + 1} R_n \left(1+z^2\right)^{-\frac{\beta+1}{2}} e^{-\gamma \tan^{-1}(z)} \left\{ n\left( z - \dfrac{\gamma}{\beta - n + 1/2} \right) P_n^{(\mu,\nu)}(iz) \right.$<br>$\left. -i \left[ \dfrac{(\beta - n + \frac{1}{2})^2 + \gamma^2}{\beta - n + 1/2} \right] P_{n-1}^{(\mu,\nu)}(iz) \right\}$ |
| **Dirac-Pöschl-Teller** | $\phi_n(r) = R_n(1-z)^{\beta/2}(1+z)^{-\gamma/2} P_n^{(\mu,\nu)}(z)$<br>$z = \cosh(2\lambda r), \beta = F/\lambda, \gamma = G/\lambda$<br>$\mu = \beta - \tfrac{1}{2}, \nu = -\gamma - \tfrac{1}{2}$ | $\theta_n(r) = \dfrac{2\alpha\lambda}{\varepsilon_n + 1} R_n (1-z)^{\frac{\beta-1}{2}} (1+z)^{-\frac{\gamma+1}{2}} \left\{ n\left( z - \dfrac{\beta + \gamma}{\beta - \gamma + 2n - 1} \right) P_n^{(\mu,\nu)}(z) \right.$<br>$\left. -2 \left[ \dfrac{(\beta + n - \frac{1}{2})(-\gamma + n - \frac{1}{2})}{\beta - \gamma + 2n - 1} \right] P_{n-1}^{(\mu,\nu)}(z) \right\}$ |

10